\documentstyle[12pt,aasms]{article}
\newcommand{\etal}{{\it et~al.}}
\newcommand{\gsim}{\lower .5ex\hbox{$\buildrel > \over {\sim}$}}
\newcommand{\lsim}{\lower .5ex\hbox{$\buildrel < \over {\sim}$}}
\newcommand{\mdot}{{\rm M}_\odot}
\newcommand{\refmark}[2]{#1}
\newcommand{\comment}[1]{}

\begin{document}

\title{A Gamma-Ray Monte Carlo Study\\
       of the Clumpy Debris of SN1987A}
\author{Adam Burrows}
\affil{Departments of Physics and Astronomy, University of Arizona, Tucson, AZ
85721}
\and
\author{Kenneth A. Van Riper}
\affil{Los Alamos National Laboratory, Transport Methods
Group, Los Alamos, NM  87545}

\begin{abstract}

We have performed Monte Carlo calculations of gamma-ray transport
in models of the clumpy debris cloud of the LMC supernova, SN1987A, to study
the influence of
composition mixing and heterogeneity on its emergent gamma-ray and X-ray
fluxes.
In particular, we have focused on the problematic Ginga band (16 -- 28 keV)
flux at day
600,  whose measured value was an order of magnitude higher than predicted by
previous theory.  We find that the hydrogen of the envelope could not have been
intimately mixed
with the heavy elements of the core and that the hydrogen/helium volume filling
factor interior
to 4000 km s$^{-1}$ must have been large ($\ge 40\%$).
Furthermore, we demonstrate that
that one can not mimic the effects of clumping by artificially decreasing the
photoelectric cross sections by some factor.  A physical separation of the
Compton scattering
region and the regions occupied by the high-Z elements is required.
The 600-day models that best fit both the line data at 847 keV and 1238 keV and
the measured Ginga band
fluxes suggest that as much as 50\% of the explosively
produced $^{56}$Ni stayed interior to 1000 km s$^{-1}$ and 2 M$_\odot$. The
$^{56}$Ni
may have been more centrally-concentrated than in the benchmark models.
$^{56}$Ni filling factors greater than 60\% are not preferred, since such
models
are too good at absorbing photons below 100 keV.
Furthermore, a total envelope mass
between 10 $\mdot$ and 15 $\mdot$ is favored.

\end{abstract}

\section{Introduction}

One of the major highlights of the campaign to observe the recent LMC
supernova,
SN1987A, was its early detection in gamma-ray lines and hard X-rays
(\cite{mat88}; \cite{san88}; \cite{coo88}; \cite{res89}; \cite{mah88};
\cite{geh88}; \cite{sun88}; \cite{ube88}; \cite{tee89}). An international
effort involving satellites (NASA's Solar Maximum
Mission and the Compton \underbar{G}amma-\underbar{R}ay
\underbar{O}bservatory, ROENTGEN on Russia's MIR-KVANT) and balloons
({\it e.g.,}\ \cite{tue90}) successfully detected the 847, 1238, 1771, and 2598
keV
gamma-ray lines of the $(^{56}$Ni$\to)^{56}$Co$\to^{56}$Fe decay sequence and
their Comptonized continuum down to a few tens of keV. In addition, the GRO
detected at day 1620 the 122 keV/136 keV blend plus continuum from the decay
of $(^{57}$Ni$\to)^{57}$Co to $^{57}$Fe, long predicted to be explosively
produced with the more abundant $^{56}$Ni (\cite{kur92}; \cite{cla69};
\cite{cla74}; \cite{col69}).  That such radioactivities were strongly indicated
by the light curves and optical/IR spectra of SN1987A and other supernovae in
no way detracts from their direct detection via gamma-ray spectroscopy.

However, as SN1987A evolved, theorists were forced by the shape of its UVOIR
light curve and the early emergence of the 847 and 1238
keV lines  (starting day 180) and continuum (starting day 130)
from $^{56}$Co decay to revise their
assumptions about the spatial
distribution of the $^{56}$Co in the debris (\cite{woo87}; \cite{geh87};
\cite{cha87}; \cite{ebi88}; \cite{mcc87}). A central point source of $^{56}$Ni
would
have been visible only after 400--600 days. Mixing of $^{56}$Ni out beyond the
Hydrogen/Helium interface to velocities of $\sim$4000 km s$^{-1}$ and interior
model
masses
greater than 10 $\mdot$ was invoked to keep pace with SN1987A's actual behavior
(\cite{pin88}; \cite{nom88}). This implied that there had been severe
hydrodynamic instabilities during the outward progress of the supernova shock
wave that mixed $^{56}$Ni into the stellar envelope from its birth at
$\sim$500 km s$^{-1}$ and $\lsim 1\ \mdot$ in the deep interior of the debris
(\cite{arn89}; \cite{her92}). Nevertheless, multi-dimensional simulations of
the Rayleigh-Taylor mixing instabilities have failed to date to reproduce the
extent of the $^{56}$Ni penetration and the detailed character of the
clumps. Instabilities during the explosion itself are suspected to be
necessary, but this has yet to be demonstrated (\cite{her92}; \cite{bf92};
\cite{bhf94}).

The models such as ``10HMM'' (\cite{pin88}) and ``sn14e1'' and ``sn11e1''
(\cite{nom88}) that
were altered to fit the optical and hard photon data had spherical
element distributions determined empirically, not physically. The isotopes of
the pre-supernova progenitor and those produced explosively were all mixed on
microscopic scales and no attempt was made to model the heterogeneity of the
ejecta. Such heterogeneity is, however, indicated by the ragged line profiles
observed in the optical and infrared (\cite{jen93}; \cite{han93};
\cite{spy90}; \cite{sta88}) and at some level in the gamma-ray
(\cite{tue90}). A reluctance to tinker further with models that seemed to fit
the gross properties of SN1987A's UVOIR, gamma-ray, and X-ray evolution is
understandable, particularly since the theoretical calculations of the
Rayleigh-Taylor instabilities are even now in an embryonic stage.

However, from day 130 through day 650 the Japanese satellite {\bf Ginga}
detected
X-rays from SN1987A in the 16--28 keV band (\cite{ta88a}; \refmark{b}{ta88b};
\refmark{c}{ta88c}; \refmark{1991}{tan91}) that have never been successfully
fit after day 520, and have never been fit after day 400 without excessive
artifice. At day 600, the discrepancy at 16--28 keV (hereafter, the ``Ginga
band'') between the predictions of models that fit the gamma-ray {\it lines}
to within a factor of two is between a factor of 5 and 25. The Ginga photons
originate from $\sim$5--20 Compton down-scatterings of
gamma-ray lines from $^{56}$Co decay, $^{57}$Co decay, and positron
annihilation (\cite{mcc93}). To emerge at these relatively low energies,
the hard photons  must not only
scatter many times before escaping, but must avoid being absorbed
photoelectrically by the multitude of heavy elements, in  particular those of
the iron group. Since the emergent Ginga photons must have followed a tortuous
path, they are ideal probes of the entire structure and composition of the
supernova debris cloud. Even better than the nuclear lines, the Ginga photons
provide a global ``X-ray'' that is diagnostic of the element yields,
distributions, and mixing. It is this great investigative leverage of the
Ginga band and the failure to date to explain the late-time Ginga flux that
has motivated this paper. Here, we distill the results of about one hundred
gamma-ray Monte Carlo calculations we have recently performed in clumpy, {\it
heterogeneous} models of SN1987A. We focus on the Ginga flux at day 600 to
investigate the degree of mixing of light (hydrogen and helium) and heavy
(Ni$-$Co$-$Fe,
Si, etc.) elements, their volume filling factors, and their radial
distributions, using as starting points the models sn14e1 and 10HMM.

\refmark{Kumagai \etal\ (1989)}{kum89} attempted to fit the late Ginga data
with a spherical model by artificially cutting the photoelectric
cross-sections down by a factor of 9. This was done to mimic the
self-shielding effect of dense clumps of intermediate to heavy
nuclei. However, their ansatz failed beyond day 520 and they
resorted to a buried pulsar, not now indicated (\cite{spd91}), to account for
the last data
points. The {\it et al.} (1990) explored spherical
models with different velocities and masses, but fared no better than
\refmark{Kumagai \etal\ (1989)}{kum89} beyond day 520. \refmark{The \etal\
(1993)}{the93} used a cloud model inspired by \refmark{Bowyer \& Field
(1969)}{bow69}, but though their paper contained many interesting suggestions,
they too failed to explain the late Ginga data. The essential problem,
discussed
by \refmark{McCray (1993)}{mcc93}, is that the Ginga photons originate in
regions of freshly-synthesized iron-group elements (with total mass$\,
\sim0.07\ \mdot$), but
must
down-scatter in regions composed predominantly of hydrogen and helium $(\ge
5\ \mdot)$. This implies that iron and hydrogen in SN1987A are physically
segregated,
and are not intimately mixed, as was assumed in models 10HMM, sn14e1, and
sn11e1.
Otherwise, if the photons were to scatter in hydrogen
significantly contaminated by heavy elements, the probability would be high
that they would be absorbed before reaching the energies of the Ginga band.
The optical/IR data obtained from SN1987A say
little about the degree of segregation of the hydrogen/helium, the
iron-peak, and the intermediate-mass elements.
However, the gamma-ray Monte Carlo simulations we present in this paper
demonstrate and quantify this effect.

Furthermore, and as reviewed in \S II, while there is optical evidence for
hydrogen in large clumps and at low
velocities and radii (\cite{han93}), these data do not constrain the total mass
of hydrogen nor
its volume filling factor.
In addition, while oxygen
is seen in many small clumps (\cite{sta88}) and its total mass may reach 1.5
$\mdot$ (\cite{chu94}), its
volume filling factor, now estimated at $\sim$10\%, is uncertain by more than a
factor of five.
In fact, it is only the product of the oxygen mass and its filling factor that
is constrained by the optical data (\cite{mcc93}).
While there is some
evidence for radioactive cobalt at small radii (\cite{han93}) and the
iron-peak elements are known to extend to 4000 km s$^{-1}$, their radial
{\it distribution} is not well-known. Finally, Li, McCray, \& Sunyaev (1993)
suggest on the basis of infrared line data
that the iron-peak elements have a
disproportionately high volume filling factor $(10\%-90\%?)$.  However, only
their lower limit is firm.
The Ginga data at day 600 can be used in conjunction with our new gamma-ray
transport calculations to address all these issues.
In particular, a comparison
between these data and the results of the Monte Carlo calculations
allows us to constrain
the hydrogen/helium mass, the hydrogen and iron filling factors, and the radial
distribution
of the $A=56$ elements.

\eject

\section{Idealized Models of the Heterogeneous Structure of the SN1987A Ejecta}

Optical and infrared data have already been used to estimate the radial
distributions and volume filling factors of the debris (\cite{mcc93}).  We were
guided by these data in manufacturing
our ejecta models and in posing the questions we wanted our study to address.
A synopsis of what they
suggest can provide the context of our studies of SN1987A heterogeneity.
Ultimately, we
want to determine whether a consistent picture can be constructed and what new
and useful insights can be gained by focusing on the Ginga band near day 600.

\refmark{Hanuschik \etal\ (1993)}{han93} have analyzed $H\alpha$
data from day 115 to day 673 and find in the jagged line profiles evidence for
$\sim 5$ large ``bumps'' in the radial velocity range
$-2000$ km s$^{-1}$$<v<1300$ km s$^{-1}$, with widths near 500 km s$^{-1}$.
They say that some
hydrogen is at quite low velocities, significantly lower than the minimum
velocity $(\sim3000$ km s$^{-1}$) hydrogen would have had if the nested
onion-skin
structure of the progenitor had not been violated. They see no evidence of
intact
shells and see peaked profiles that can also imply deep penetration. They
think they see ``extra emission'' in the bumps, symptomatic of proximity to
some of the radioactive source ($^{56}$Co) of Compton electrons and gamma
rays. \refmark{Hanuschik \etal }{han93} can not say much about the H/He
volume filling factor or the total hydrogen mass and are unable to say
anything about the possible existence of ``fingers.''

\refmark{Cannon \& Stathakis (1988)}{sta88}
observe at high spectral resolution
$({\lambda\over\Delta\lambda}\sim 30000)$ the 6300, 6363\AA\ [OI] doublet and
identify many (they say $\sim 50$) small-scale oxygen clumps down to widths of
$\Delta v\sim 60$ km s$^{-1}$. \refmark{Chugai (1994)}{chu94} interprets these
same
data with a statistical model of $\sim 2000$ clumps of ``oxygen'' with an
average clump ``radius'' of 60 km s$^{-1}$, a clump mass of $\sim 10^{-3}\
\mdot$,
a total mass of 1.2--1.5 $\mdot$, and a volume filling factor of 10\%. His
model puts all the emitting oxygen at radial velocities below $\sim$1700 km
s$^{-1}$
(at half the radius of the outer ``nickel'' bullets). \refmark{Wampler
(1994)}{wam93} lends indirect support to the larger oxygen clump number with a
super-resolution $({\lambda\over\Delta\lambda}\sim 10^6)$ [OI] spectrum in
which clumpy structure is
seen below $\Delta v\sim 60$ km s$^{-1}$. No Fourier or wavelet analysis of
these
data has been circulated.

\refmark{Li, McCray, \& Sunyaev (1993)}{li93} have analyzed the nickel,
cobalt, and iron data in the near- and mid-infrared and have constructed a
model for the iron-peak spatial distribution. They see $\sim$60--100 clumps
of Ni-Co-Fe below 2500 km s$^{-1}$, with a volume filling factor greater than
10\%. They infer a ``frothy'' structure of low-density ``iron'' surrounded by
higher density filaments of H, He, and other elements. The evidence for high
density hydrogen and helium is weak, but the high ``iron'' filling factor is
consistent
with the predicted expansion of the radioactive nickel (``nickel bubble'') and
cobalt during the first weeks of explosion (\cite{bas94}). (Note that the
freshly
synthesized iron-peak nuclei comprise only about 1\% of the ejectum mass.)

Clearly, an infinite number of parameters could characterize
the heterogeneity of clumpy supernova debris.  It might at first glance seem
natural to use
the published  (\cite{arn89}; \cite{her92}) two-dimensional
hydrodynamic simulations of SN1987A's envelope for our Monte Carlo
calculations. However,  since
$^{56}$Ni penetrates to only 2000 km s$^{-1}$ in those
simulations and they do not credibly include the effects of gamma-ray and
heat leakage at late times,
their use was not advisable. Without a specific model to falsify, we were
forced to construct ``artificial'' debris clouds
with gross properties that were still consistent
with most of what was known. The major constraints on these models were the
given isotope masses of the
original progenitor models (10HMM or sn14e1), the observed velocity
range of $^{56}$Ni, and the homology of the velocity field.
To make this project manageable,
we focused on a set of specific synthetic structures and element
distributions (with variations) that we felt would bracket the range
of realistic ejecta. Our models spanned a far broader range of density
contrasts
and clump size distributions than are to be found in the two-dimensional
hydrodynamic simulations of SN1987A.  We also looked at the effect of various
degrees of central
concentration of both iron-peak elements (hereafter, composition 2)
and intermediate mass elements
($2 < Z < 26$, hereafter, composition 0). Identifying the matter of the H/He
zones of the
original stellar models with composition 1, we distributed the three
compositions in two or three phases. (For the two-phase models, composition 0
included helium and composition 1 was pure hydrogen.) A phase was either a
denser radial
finger (various total finger volumes, and hence densities, were used)
or the interstitial space between the
fingers. The two-phase calculations involved one set of denser fingers
surrounded by lower-density interstitial gas and the three-phase calculations
involved two set of fingers of different compositions surrounded by an
interstitial phase of a third composition. In the two-phase calculations, the
fingers or the space between the fingers could be made of different
mixtures of compositions
(composition 2 with composition 1, composition 2 with composition 0, etc.).
Breaking the ejecta
thusly into phases was suggested by the ``onion skin'' structure of supernova
progenitors and
the hydrodynamic simulations of \cite{arn89} and  \cite{her92}.

A set of planes and cones were introduced to partition the initial spherical
model and define the finger regions.  The planes are parallel to and pass
through a single axis
of the star
and are equally spaced in the meridional angle, $\theta$, around this axis.
This
line is also the axis of the cones.  The double-sheeted cones share a common
vertex at the center of the star.
The polar angles $\phi$ of the cones are chosen to
give equal volumes between successive cones, including the interior of
the polar-most cone and between the lowest cone and the equator.
Any plane through the axis is a plane of reflection symmetry, as
is the equator.  We label the angular division by the number of planes
$N_\theta$ and cones $N_\phi$.  There are  $2 N_\theta \times 2(N_\phi+1)$
equal volume angular elements.  Each element is assigned composition 0, 1,
or 2 (or some mixture in the two-phase model); these assignments are
the same for each radial shell.  Each composition has a single density
within a shell.  When the matter of a composition is
contained in a sparse set of isolated angular elements, dense fingers
are formed.  It should be kept in mind when considering these models that the
detailed shape of a Rayleigh-Taylor
plume ({\it i.e.,} whether it is square or rounded) can not be discerned in
emergent hard spectra.
Since diffusing photons quickly lose their orientation as they scatter,
the output of a transport calculation depends more on the integral, average,
and global
properties of the debris cloud than on its details.  Models with a variety of
``tilings,'' for example with and without
channels to the surface, give roughly the same answers for the same global
parameters ({\it i.e.},
filling factors, nickel mass, density contrasts, degree of central
concentration).
It is in this spirit that we constructed the models we did. Furthermore,
calculations with this relatively simple geometry are
more efficient than those with geometries that are more elaborate.
With these models we have spanned
in a systematic way a very broad range of volume filling factors, density
contrasts, and nickel distributions.

Figure 1 depicts a typical two-phase finger pattern and Figure 2 shows
a three-phase geometry.  We ran a large number of models to investigate various
element mixtures,
finger densities, finger numbers, and the angular distribution of fingers.
The three-phase models were parametrized by the volume filling factors of
the compositions, which are determined by the finger patterns.
These filling factors, together with the starting spherical model, the fixed
element
masses, and the radial extent of the fingers implicitly
give the finger density contrasts. Finger contrast ratios from one to almost
600 were studied. In all, the gamma transport in more than sixty models at day
600 was calculated. The almost thirty models in Table 1 summarize this more
extensive list. (We also did a total of more than fifty Monte Carlo
calculations at days 200, 400, 500, and 700.)

We used both a coarse ($N_\theta\times N_\phi = 8\times8$) and a fine
($N_\theta\times N_\phi = 24\times 24$) angular division in both the two- and
three-phase models.
Tables 2a and 2b show examples of the angular assignment
patterns.
In some cases, a three-phase template was created by inserting fingers of a
third composition at random positions in a two-phase pattern. Table 1 lists the
model
names, but not the logic of the character strings that define them. The first
field (each field is separated by a period) indicates the original SN1987A
model used (either sn14 (sn14e1) or hmm (10HMM)). The letters after the
original model designator in the first field indicate special radial
composition distributions. The letter ``e'' means that all the composition
fractions were
constant from the center to 3000 km s$^{-1}$, ``f'' means that the composition
fractions were flat at their original central SN1987A model values, and ``c''
means that the $^{56}$Ni distribution was constant out to 7 $\mdot$. The
latter represents a significant central concentration of $^{56}$Ni, since it
has been inferred (and the original models assume) that some nickel bullets
penetrated to at least 12 $\mdot$. When no letter is given, or unless otherwise
indicated, the original (default) model radial element distributions should be
assumed. The next field (always .600 for the models considered here)
identifies the epoch (in days) of SN1987A. The digits in the last
field indicate the angular division (e.g. $8\times 8$,
$24\times 24$). The terminal letter identifies the combination of filling
factors, number of phases, and additional special characteristics for that
specific model. Table 3 depicts the meanings of these terminal letter
designators for the models listed in Table 1 and should be considered an
important footnote to it.

\section{The Monte Carlo Code}

To calculate hard photon transport in spherical supernovae, we had formerly
constructed a Monte Carlo code based on the variance-reduction algorithm of
\refmark{Pozdnyakov, Sobol, \& Sunyaev (1983)}{poz83} (\cite{the90}). Rather
than redesign that code to handle arbitrary geometries, we decided to employ
the MCNP\footnote{MCNP is a trademark
of the Regents  of the University of California, Los Alamos National
Laboratory.}
code developed and maintained by the Transport Methods Group
of Los Alamos National Laboratory (\cite{bri86}; \cite{for90}).
MCNP is a general-purpose Monte Carlo code
for calculating the time-dependent continuous energy transport of neutrons,
photons, and/or electrons in three-dimensional geometries.
The code has been successfully compared against numerous benchmark problems
and is subject to formalized quality assurance procedures.
MCNP is used for many applications, including nuclear oil well
logging, medical imaging and radiotherapy, nuclear reactors (both fission and
fusion), space science, nuclear safeguards, personnel dosimetry, detector
design and analysis, and radiation shielding.
We have used MCNP in previous studies of gamma-rays from Type Ia supernovae
(\cite{bsvr91}) and found excellent agreement with the results of other Type Ia
calculations.
(MCNP does not account for Doppler shifts in photon creation and
interactions and thus does not give realistic line shapes; the flux in
a line and the continuum spectrum are unaffected by this neglect.)
The photon physics model includes Compton scattering, photoelectric absorption,
fluorescence, and the Auger effect.
Although MCNP is rich in variance-reduction techniques, we used
only importance splitting based on radius.
For our calculations, a continuous (rather than multigroup) energy
representation
was used for the photons.

MCNP can
follow the transport of secondary Compton electrons and their bremsstrahlung
photons. Bremsstrahlung had originally seemed to us a good source of Ginga
photons, but we later discovered that above 15 keV, bremsstrahlung does not
contribute a competitive fraction of the flux. Compton photons still dominate
down to the Ginga band. Nevertheless, most of the calculations we report here
were done with the bremsstrahlung yield included.  The bremsstrahlung is
calculated in the ``thick-target'' model, where the radiation distance of
the electron is assumed to be much less than other scales and the
spatial transport of the electron away from the emission site is neglected.
A similar bremsstrahlung treatment was used by
Clayton \& The (1991).

As in our original spherical supernova studies, the $^{57}$Ni and $^{56}$Ni
decay schemes, energies and branching ratios were taken from \refmark{Browne
\etal\ (1978)}{bro78} and it was assumed that all positrons form positronium
before annihilating (\cite{bus79}). MCNP comes with its own photoelectric
cross section library. Between $10^5$ and $5\times 10^5$ decays were followed
per run,
and for each run a statistical error estimate was made. The Ginga fluxes that
we quote have a statistical error of between 0.5 and 3 percent.

\section{The Ginga Band Model Fluxes at Day 600 and Their Interpretation}

Figure 3 shows the Ginga band data accumulated during the SN1987A campaign
(\cite{ta88a}; \refmark{b}{ta88b}; \refmark{c}{ta88c};
\refmark{1989}{tan89}). Superposed are the 600-day predictions of a few
representative models from Table 1. As the figure shows, the 16--28 keV Ginga
band flux was measured to be $4\pm 2\times 10^{-4}$ photons cm$^{-2}$s$^{-1}$
between days 550 and 650. The fact that the Ginga fluxes of the spherical
models
sn14.600.sph and hmm.600.sph at day 600 (Table 1) are only $6.96\times 10^{-5}$
 cm$^{-2}$s$^{-1}$ and
$1.61\times 10^{-5}$ cm$^{-2}$s$^{-1}$, respectively, is the ``Ginga Problem.''
The former is
lower by a factor of $\sim 6$; the latter by a factor of $\sim25$. The
sn14e1-derived model has four times the flux of the 10HMM-derived model
predominantly because sn14e1 has a larger mass of hydrogen on which to
Compton-scatter (5.6 $\mdot$ versus 4.5 $\mdot$), has a larger fraction of its
$^{56}$Ni near the center, and has less than half the burden of silicon,
sulfur, argon, and calcium isotopes (respectable photoelectric absorbers). All
these factors have a bearing on what fits.

Table 1 (with reference to Table 3) contains the major results of our Monte
Carlo simulations and should be scrutinized closely. We summarize here what it
says. By comparing sn14.600.2424h$'$ with sn14.600.2424k we can see the
drastic effect of mixing hydrogen with the iron-peak
nuclei. The ``h$'$'' model flux is {\it six} times lower. A
similar effect of intimate mixing is seen in model sn14.600.2424g. In fact,
none of the models that mixed compositions 2 and 1 came anywhere near the
data. From this, we conclude that most of the $^{56}$Ni and H/He could not
have mixed and that these compositions {\it must} be separated in SN1987A.

A comparison of models hmm.600.88h and hmm.600.2424h shows that a finer finger
division with larger density contrasts yields a larger Ginga flux, all else
being equal. The ``2424'' model has more than twice the Ginga flux of the
``88'' model. This is as one may have anticipated and is a universal feature
of the models in Table 1. The generically small effect of including the
bremsstrahlung of Compton electrons can be discerned by comparing models
hmm.600.88h and hmm.600.88hnob. Bremsstrahlung increases the Ginga band flux
at day 600 by no more than 2--3\%. The effect of clumping all the way to the
surface (instead of to 4000--5000 km s$^{-1}$) can be seen by comparing models
sn14.600.88t and sn14.600.88r. Having channels to the surface (not expected or
indicated in the early optical data) increases the escape probability of the
Ginga photons by only 7--10\%. At day 600, the outer envelope is generally
transparent and it is in the inner zones interior to 3000 km s$^{-1}$ that the
Compton X-rays are created (\cite{the90}).

We now turn to the effect of volume filling factors (f) on the emergent Ginga
flux. (Refer as needed to Tables 1 and 3.) Models ``t'' have $f(0)=0.819$,
$f(1)=0.083$ and $f(2)=0.097$. Most of
the volume in the clumped region is filled with low-density intermediate Z
elements. Models ``u'' and ``v'' have $f(1)=0.72$ and 0.79, respectively, and
each has $f(0)=0.097$ (only $\sim$10\%). Otherwise, they are
similar. Comparing sn14.600.88t and sn14e.600.88t with models sn14.600.88u,
sn14.600.88v, sn14e.600.88u, and sn14e.600.88v shows the dramatic importance
of a large H/He volume filling factor. It is hard with a small $f(1)$
($<20\%$) to fit the Ginga data at day 600. All the models with the highest
fluxes have large $f(1)$ $(>40\%)$. As a
comparison of sn14.600.88s $(f(2)=0.097)$ with sn14.600.88p $(f(2)=0.417)$
shows, it is more
nearly the {\it sum} of the composition 0 and composition 2 volume filling
factors that is constrained. This is also implied by the high flux of model
sn14.600.2424k, for which all $Z > 2$ elements are contained in very dense
fingers.
This indicates that most of the H/He (composition
1) is not contaminated by the strong photoelectric absorbers of compositions 0
and 2, that a high $f(1)$ is preferred, and that $f(0)$ and $f(2)$ are not
individually determined by the Ginga data.

One tentative conclusion of this work is that despite the enhancements in the
Ginga flux from the modifications described above, central
concentration of the iron peak elements beyond that provided by the fiducial
SN1987A
models can further improve the fits. Placing the ``nickel'' deeper
inside the debris increases the Compton opacity and allows more Compton
photons to be created. At day 600, nickel in the outer envelope creates gamma
line photons that escape uselessly to infinity. This effect is amplified in
the ``e'' and ``f'' series calculations that put too much $^{56}$Ni, $^{56}$Co,
and $^{56}$Fe at large
radii.  We see the effect of central concentration by comparing model
sn14c.600.2424s with model sn14.600.2424s or model sn14c.600.88u with model
sn14.600.88u. The ``c'' series has 50\% to 70\% higher Ginga fluxes.

Importantly,
as indicated in the full spectra depicted in Figures 4 and 5, central
concentration tends to decrease the emergent {\it line} fluxes in the MeV
range. The theoretical models have generally overestimated the late ($>$300d)
line fluxes (\cite{the90} and their Figures 3 and 4). This mismatch is only
about
a factor of two and, given the low S/N ratio of gamma line data, was not
considered a major problem. Our calculations indicate that while concentrating
more ``nickel'' in the center may contribute to the solution of the Ginga
Problem, it can also
resolve the milder 847 and 1238 keV line flux discrepancies. In addition,
concentrating the nickel much further than we have done in this study cuts
the line fluxes at day 614 by {\it too} much (\cite{tue90}). Thus, and very
approximately, we have both lower
and upper limits
on the nickel radial
distribution. Centrally-concentrated model sn14c.600.2424s fits all the data to
within 1$\sigma$.
This implies that while some
of the explosively produced $^{56}$Ni was obviously flung to high radii,
velocities, and interior masses, a large portion of it may not have been. This
result
may be of importance in understanding the hydrodynamic instabilities that
violated the shell structure of the progenitor of SN1987A. The two-phase model,
sn14.600.2424k,
yields the highest Ginga band flux of the set depicted in Table 1, but does so
with a density
ratio between the $Z > 2$ and the hydrogen/helium ``phases''$\,$of close to
600.  This ratio is perhaps
too extreme to be physical, but its success in fitting the Ginga band at day
600 should not
be ignored.

A final result of our series of Monte Carlo calculations is the recognition
that the
debris of SN1987A had to have at least 5 $\mdot$ and preferably closer to 10
$\mdot$ of hydrogen and helium to give any Ginga flux at all at day
600. In fact, a model with only 3 $\mdot$ of hydrogen would miss the observed
600-day Ginga band flux by a
factor of as much as one hundred. This means that the mere detection from
SN1987A at day 600 of Ginga band photons implies that its debris mass had to be
above $\sim$7 $\mdot$. This result is
consistent with the large sn14e1 and 10HMM model envelope masses.

\section{Conclusions}

Since the photons that eventually emerge from SN1987A in the Ginga band have
scattered numerous times
and have lost their orientation,
artificial debris models with a range of filling factors, average density
contrasts, and integral properties
can be used in conjuction with the observed Ginga band photons to constrain
these global properties.  The detection by Ginga on day 600
of photons in the 16--28 keV band at the substantial level
published implies that in SN1987A the hydrogen could not have been intimately
mixed with
either the iron-peak elements or the elements with $Z$'s between 2 and 26.
In addition, the hydrogen/helium volume filling factor interior to 4000 km
s$^{-1}$ must have
been large $(\gsim\ 40\%)$.  Furthermore, perhaps as much as 50\% of the
explosively
produced $^{56}$Ni stayed interior to 1000 km s$^{-1}$ and 2 $\mdot$; the
$^{56}$Ni may be
more centrally concentrated than in the published benchmark SN1987A
models.  We have demonstrated that one can not mimic the effects
of clumping by simply decreasing the effective photoelectric cross sections by
some factor ({\it cf.} \cite{kum89}). A physical segregation of the Compton
scattering
region from the regions of the ``photoelectric'' elements is required.
Our best
heterogeneous models not only solve the Ginga Problem at day 600, but
improve the fits of the 847 and 1238 keV line data with
theory. A large $^{56}$Ni filling factor ($f(2) > 60\%$) is disfavored, since
such models
are too good at absorbing photons below 100 keV ({\it cf.} \cite{li93}).
In fact, it is the sum of the filling factors for all of the $Z>2$ elements
that is probed by the Ginga data and a sum value less than 50\% is
suggested, though a model with $f(1)=50\%$ and $f(0)+f(2)=50\%$ can be
accommodated by the data.
No model with a total envelope mass less than 5 $\mdot$ can
possibly fit all the line and Ginga band data between days 300 and 650. A
total envelope mass between 10 $\mdot$ and 15 $\mdot$ is favored.

\section{Acknowledgments}

The authors would like to thank Lih Sin The, Anurag Shankar, Phil Pinto, and
Dick McCray for comments and input. This project could not easily have been
completed without the MCNP code structure, for which we gratefully acknowledge
the Los Alamos National Laboratory. Adam Burrows would like to thank NASA and
the NSF for providing crucial financial support under grants
NAGW-2145 and AST92-17322, respectively, during the execution of these
calculations. Further information on MCNP can be found at World-Wide-Web
address
http://www-xdiv.lanl.gov/x6/mcnp/mcnp.html.

\vfill\eject

\vbox{
\centerline{{\bf Figure Captions}}
\baselineskip=20pt
\setcounter{page}{26}
\noindent Figure 1: Two-Phase pattern.  Three-dimensional representation of the
two-phase
pattern pat\_88\_10.  The fingers are shaded and the region between the fingers
is transparent.  The innermost zone is opaque.  The outermost zone contained
within the
fingers is shown in a darker shade as an aid to the eye. The fingers encompass
10\% of the volume.
\smallskip

\noindent Figure 2: Three-Phase pattern. Three-dimensional representation of
the
three-phase patterns pat\_2424\_tw and pat\_2424\_tx.  In pattern
pat\_2424\_tw, the compact
fingers containing composition 2 are dark grey, composition 0 is light grey,
and
composition 1 is transparent.  For pat\_2424\_tx, the composition assignments
are 0 in the
dark grey, 2 in light grey, and composition 1 is again transparent.  The
outermost
radial zone of the light grey volume has been made transparent to permit better
recognition of the dark fingers. The dark fingers represent 0.83\% of the
volume,
the light regions 43.7\% of the volume, and the transparent composition 1
regions 55.5\% of the
volume.

\noindent Figure 3: A plot of the measured flux in the Ginga band (16 - 28 keV)
versus time.
The error bars are 1-sigma error bars. The solid line is the theoretical
prediction for the
10HMM model of Pinto \& Woosley (1988) and the dashed line is the theoretical
prediction
for the sn14e1 model of Nomoto \etal\ (1988), both as calculated by The \etal\
(1990).
This paper focuses on the discrepancy at 600 days. The symbols at day 600 are
representative
results from our multi-dimensional Monte Carlo calculations. The models in
order of increasing flux
at 600 days are hmm.600.sph (B), sn14.600.2424h$'$ (J), sn14.600.sph (A),
sn14.600.88t (S), sn14.600.88p (M),
sn14.600.2424s (R), sn14c.600.2424s (W), and sn14.600.2424k (L).  The capital
letters used to
identify the models here
are the same letters used in the listing on Table 1.

\noindent Figure 4: A superposition of the spectra (in photons
cm$^{-2}$s$^{-1}$keV$^{-1}$) from 10 keV to 4 MeV at day 600 for models
hmm.600.sph (B), sn14.600.2424h$'$ (J), sn14.600.2424g (K), sn14.600.sph (A),
sn14.600.2424s (R), and sn14.600.2424k (L).
The spikes are the lines and their heights are the line fluxes in
photons/$[cm^2\cdot s]$.
The capital letters used to
identify the curves here
are the same letters used in the listing on Table 1.

\noindent Figure 5: Same as Figure 4, but for models sn14e.600.88v (V),
sn14.600.88t (S), sn14.600.88p (M),
sn14.600.88v (P), and sn14c.600.2424s (W).
}


\begin{thebibliography}{}
\bibitem[Arnett, M\"uller, \& Fryxell 1989]{arn89}
\reference Arnett, W. D., Fryxell, B., \& M\"uller, E., 1989,
   Ap. J. (Letters), 341, L63
\bibitem[Basko, 1994]{bas94}
\reference Basko, M. 1994, Ap. J. 425, 264
\bibitem[Bowyer \& Field 1969]{bow69}
\reference Bowyer, C. \& Field, G. 1969, Nature, 223, 573
\bibitem[Briesmeister 1986]{bri86}
\reference Briesmeister, J. F., Editor  1986, {\it MCNP--A General Monte Carlo
Code for Neutron and Photon Transport, Version 3A}, Los Alamos National
Laboratory Report LA--7396--M, Rev. 2
\bibitem[Browne \etal\ 1978]{bro78}
\reference Browne, E., Dairiki, J. M.,
Doebler, R. E., Shihab-Edin, A. A., Jardine, L. L., Tuli, J. K., \& Buyrn,
A. B. 1978, in {\it Table of Isotopes}, ed. C. M. Lederer \& V. S. Shirley
(New York: Wiley).
\bibitem[Burrows \& Fryxell 1992]{bf92}
\reference Burrows, A. \& Fryxell, B.  1992, {\it Science}, 258, 430
\bibitem[Burrows, Hayes, \& Fryxell 1994]{bhf94}
\reference Burrows, A., Hayes, J., \& Fryxell, B. 1995, Ap. J., in press
\bibitem[Burrows, Shankar, \& Van Riper 1991]{bsvr91}
\reference Burrows, A., Shankar, A., \& Van Riper, K. A. 1991, Ap. J.
(Letters), 379, L7
\bibitem[Bussard, Ramaty, \& Drachman 1979]{bus79}
\reference Bussard, R. W., Ramaty, R., \& Drachman, R. 1979, Ap. J. 228, 928
\bibitem[Cannon \& Stathakis 1988]{sta88}
\reference Cannon, R. D. \& Stathakis, R. A., 1988, Anglo-Australian
Observatory
Newsletter no. 45.
\bibitem[Chan \& Lingenfelter 1987]{cha87}
\reference Chan, K. W. \& Lingenfelter, R. E., l987, Ap. J. (Letters), 318,
L51
\bibitem[Chugai 1994]{chu94}
\reference Chugai, N. N., 1994, Ap. J. (Letters), 428, L17
\bibitem[Clayton \etal\ 1969]{cla69}
\reference Clayton, D. D., Colgate, S. A., \& Fishman, G. 1969, Ap. J., 155,
75
\bibitem[Clayton \etal\ 1974]{cla74}
\reference Clayton, D. D. 1974, Ap. J., 188, 155
\bibitem[Clayton \& The]{clathe91}
\reference Clayton, D. D. \& The, L.-S. 1991, Ap. J., 375, 221
\bibitem[Colgate \& McKee 1969]{col69}
\reference Colgate, S. A. \& McKee, C. 1969, Ap. J., 157, 623
\bibitem[Cook \etal\ 1988]{coo88}
\reference Cook, W. R., Palmer, D. M., Prince, T. A., Schindler, S., Starr,
C. H., \& Stone, E. C. 1988, Ap. J. (Letters), 334, L87
\bibitem[Ebisuzaki \& Shibazaki 1988]{ebi88}
\reference Ebisuzaki, T., \& Shibazaki, N., 1988, Ap. J., 328, 699
\bibitem[Forester \etal\ 1990]{for90}
\reference Forester, R. A., Little, R. C., Briesmeister, J. F., \& Hendricks,
J. S. 1990,
{\it IEEE Transactions on Nuc. Sci.},  37, 1378
\bibitem[Gehrels, Leventhal, \& MacCallum 1988]{geh88}
\reference Gehrels, N., Leventhal, M., MacCallum, C. J., l987, Ap. J., 322,
215
\bibitem[Gehrels, MacCallum, \& Leventhal 1987]{geh87}
\reference Gehrels, N., MacCallum, C. J. \& Leventhal, M., 1987, Ap. J.
        (Letters), 320, L19
\bibitem[Hanuschik \etal\ 1993]{han93}
\reference Hanuschik, R., \etal\ 1993 MNRAS 261, 909
\bibitem[Herant \& Benz 1992]{her92}
\reference Herant, M. \& Benz, W. 1992 Ap. J. 387, 294
\bibitem[Jennings \etal\ 1993]{jen93}
\reference Jennings, D. E. \etal\ 1993, Ap. J., 408, 277
\bibitem[Kumagai \etal\ 1989]{kum89}
\reference Kumagai, S., Shigeyama, T., Nomoto, K., Itoh, M. Nishimura, J.,
        \& Tsuruta, S., l989, Ap. J., 345, 412
\bibitem[Kurfess \etal\ 1992]{kur92}
\reference Kurfess, J. D. \etal\ 1992, Nature, 399, L137
\bibitem[Li, McCray, \& Sunyaev 1993]{li93}
\reference Li, H., McCray, R., \& Sunyaev, R. 1993, Ap. J. 419, 824
\bibitem[Mahoney \etal\ 1988]{mah88}
\reference Mahoney, W. A., Varnell, L. S., Jacobson, A. S., Ling, J. C.,
        Radocinski, R. G., \& Wheaton, WA., 1988, Ap. J. (Letters),
        334, L81
\bibitem[Matz \etal\ 1988]{mat88}
\reference Matz, S. M., Share, G. H., Leising, M. D., Chupp, E. L.,
        Vestrand, W. T., Purcell, W. R., Strickman, M. S., \& Reppin, C.,
        l988, Nature, 331, 416
\bibitem[McCray, Shull, \& Sutherland 1987]{mcc87}
\reference McCray, R., Shull, J. M., \& Sutherland, P., l987, Ap. J.
        (Letters), 317, L69
\bibitem[McCray 1993]{mcc93}
\reference McCray, R. 1993, ARAA, 31, 175
\bibitem[Nomoto \etal\ 1988]{nom88}
\reference Nomoto, K., Shigeyama, T., Kumagai, S., Itoh, M., Nishimura, J.,
        Hashimoto, M., Saio, H., \& Kato, M., 1988, in Physics of Neutron
        Stars and Black Holes, ed. Y. Tanaka (Tokyo University Press),
        p. 441
\bibitem[Pinto \& Woosley 1988]{pin88}
\reference Pinto, P. A. \& Woosley, S. E., l988, Ap. J., 329, 820
\bibitem[Pozdnyakov, Sobol, \& Sunyaev 1983]{poz83}
\reference Pozdnyakov, L. A. Sobol, I. M., \& Sunyaev, R. A. 1983, Ap. Space
Phys. Rev., 2, 189
\bibitem[Rester \etal\ 1989]{res89}
\reference Rester, A. C., Coldwell, R. L., Dunnam, F. E., Eichhorn, G.,
        Trombka, J. I., Starr, R., \& Lasche, G. P., l989, Ap. J.
        (Letters), 342, L71
\bibitem[Sandie \etal\ 1988]{san88}
\reference Sandie, W. G., Nakano, G. H., Chase, Jr. L. F., Fishman, G. J.,
        Meegan, C. A., Wilson, R. B., Paciesas, W. S., \& Lasche, G. P.,
        l988, Ap. J. (Letters), 334, L91
\bibitem[Spyromilio, Meikle, \& Allen 1990]{spy90}
\reference Spyromilio, J., Meikle, W. P. S. \& Allen, D. A., 1990,
MNRAS, 242, 669
\bibitem[Suntzeff \etal\ 1991]{spd91}
\reference Suntzeff, N.B., \etal\, 1991, Astron. J., 102, 1118
\bibitem[Sunyaev \etal\ 1988]{sun88}
\reference Sunyaev, R., \etal\, l988, Sov. Astron. Lett., 14, 579
\bibitem[Tanaka 1991]{tan91}
\reference Tanaka, Y., l991, in the Proceedings of the Xth Santa Cruz
        Workshop, "Supernovae," ed. S. E. Woosley (Springer-Verlag), p. 269
\bibitem[Tanaka 1988a]{ta88a}
\reference Tanaka, Y., l988a, in IAU Colloquium 108, Atmospheric Diagnostics
        of Stellar Evolution, ed. K. Nomoto (Springer-Verlag), p. 499
\bibitem[Tanaka 1988b]{ta88b}
\reference Tanaka, Y., l988b, in Proc. of the George Mason Workshop, "Supernova
        l987A in the Large Magellanic Cloud," ed. M. Kafatos \& A.
        Michalitsianos (Cambridge U. Press) p. 349
\bibitem[Tanaka 1988c]{ta88c}
\reference Tanaka, Y., l988c, in Physics of Neutron Stars and Black Holes,
        ed., Y. Tanaka, (Tokyo University Press), p. 431
\bibitem[Teegarden \etal\ 1989]{tee89}
\reference Teegarden, B. J., Barthelmy, S. D., Gehrels, N., Tueller, J.,
        Leventhal, M., \& MacCallum, C. J., l989, Nature, 339, 122
\bibitem[The, Burrows, \& Bussard 1990]{the90}
\reference The, L.-S., Burrows, A., \& Bussard, R. W., 1990, Ap. J., 352, 731
\bibitem[The \etal\ 1993]{the93}
\reference The, L.-S., Hartmann, D. H., Clayton, D., \& Leising, M. 1993, in
AIP Conference Proceedings \# 280, ed. M. Friedlander, N. Gehrels, \&
D. J. Macomb, p. 149
\bibitem[Tueller \etal\ 1990]{tue90}
\reference Tueller, J., Barthelmy, S., Gehrels, N., Teegarden, B. J.,
Leventhal, M. \& MacCallum, C. J., 1990, Ap. J. (Letters), 351, L41
\bibitem[Ubertini \etal\ 1988]{ube88}
\reference Ubertini, P., Bazzano, A., Sood, R., Staubert, R., Sumner, T. J.,
        \& Frye, G., 1989, Ap. J., 337, L19
\bibitem[Wampler (1993)]{wam93}
\reference Wampler, J. 1993, private communication
\bibitem[Woosley \etal\ 1987]{woo87}
\reference Woosley, S. E., Pinto, P. A., Martin, P., \& Weaver, T. A.,
        l987, Ap. J., 318, 664
\end{thebibliography}
\end{document}


\begin{center}
\begin{tabular}{lll}
\hline
\multicolumn{3}{c}{Table 1.}\\
\hline\hline
\multicolumn{1}{c}{Model Name}	&\multicolumn{2}{c}{Ginga Band Flux}\\
\noalign{\vskip-8pt}
				&\multicolumn{2}{c}{(\#/$cm^2\cdot s$)}\\
\hline
\multicolumn{3}{c}{Spherical}\\
\noalign{\vskip-4pt}
A.\ \ sn14.600.sph\tablenotemark{*}	&\hspace{2pc}	&{\bf 6.96}\mbox{\boldmath
						$\times$}{\bf 10}$^{\bf -5}$\\
\noalign{\vskip-9pt}
B.\ \ hmm.600.sph\tablenotemark{**}	&	&{\bf 1.61}\mbox{\boldmath
					$\times$}{\bf 10}$^{\bf -5}$\\
\noalign{\vskip-9pt}
C.\ \ sn14e.600.sph\hspace{6pc}	&	&$1.41\times 10^{-5}$\\
\noalign{\vskip-9pt}
D.\ \ sn14f.600.sph		&	&$1.21\times 10^{-5}$\\
\noalign{\vskip-9pt}
E.\ \ sn14c.600.sph		&	&$9.46\times 10^{-5}$\\
\multicolumn{3}{c}{Two-Phase}\\
\noalign{\vskip-4pt}
F.\ \ sn14.600.88h		&	&$1.90\times 10^{-4}$\\
\noalign{\vskip-9pt}
G.\ \ hmm.600.88h		&	&$8.30\times 10^{-5}$\\
\noalign{\vskip-9pt}
H.\ \ hmm.600.88hnob\tablenotemark{***}&	&$8.14\times 10^{-5}$\\
\noalign{\vskip-9pt}
I.\ \ hmm.600.2424h		&	&$1.87\times 10^{-4}$\\
\noalign{\vskip-9pt}
J.\ \ sn14.600.2424h$'$	&	&$4.54\times 10^{-5}$\\
\noalign{\vskip-9pt}
K.\ \ sn14.600.2424g		&	&$5.51\times 10^{-5}$\\
\noalign{\vskip-9pt}
L.\ \ sn14.600.2424k		&	&{\bf 2.60}\mbox{\boldmath
				$\times$}{\bf 10}$^{\bf -4}$\\
\multicolumn{3}{c}{Three-Phase}\\
\noalign{\vskip-4pt}
M.\ \ sn14.600.88p		&	&$1.19\times 10^{-4}$\\
\noalign{\vskip-9pt}
N.\ \ hmm.600.88p		&	&$5.10\times 10^{-5}$\\
\noalign{\vskip-9pt}
O.\ \ sn14.600.88u		&	&$1.30\times 10^{-4}$\\
\noalign{\vskip-9pt}
P.\ \ sn14.600.88v		&	&$1.46\times 10^{-4}$\\
\noalign{\vskip-9pt}
Q.\ \ sn14.600.88s		&	&$1.16\times 10^{-4}$\\
\noalign{\vskip-9pt}
R.\ \ sn14.600.2424s		&	&$1.48\times 10^{-4}$\\
\noalign{\vskip-9pt}
S.\ \ sn14.600.88t		&	&$9.15\times 10^{-5}$\\
\noalign{\vskip-9pt}
T.\ \ sn14.600.88r		&	&$9.82\times 10^{-5}$\\
\noalign{\vskip-9pt}
U.\ \ sn14.600.88q		&	&$1.22\times 10^{-4}$\\
\multicolumn{3}{c}{Fractions Flat to 3000 km/s}\\
\noalign{\vskip-4pt}
\quad\  \  sn14e.600.88t	&	&$3.81\times 10^{-5}$\\
\noalign{\vskip-9pt}
\quad\ \   sn14e.600.88u		&	&$7.60\times 10^{-5}$\\
\noalign{\vskip-9pt}
V.\ \ sn14e.600.88v		&	&$8.45\times 10^{-5}$\\
\multicolumn{3}{c}{Central, Non-Spherical}\\
\noalign{\vskip-4pt}
W.\ \ sn14c.600.2424s		&	&{\bf 2.49}\mbox{\boldmath
				$\times$}{\bf 10}$^{\bf -4}$\\
\noalign{\vskip-9pt}
X.\ \ sn14c.600.88t		&	&$1.56\times 10^{-4}$\\
\noalign{\vskip-9pt}
Y.\ \ sn14c.600.88u		&	&$1.95\times 10^{-4}$\\
\noalign{\vskip-9pt}
Z.\  \ sn14c.600.88v		&	&$2.08\times 10^{-4}$\\
\hline
\multicolumn{3}{l}{*sn14e1 default element distribution}\\
\noalign{\vskip-9pt}
\multicolumn{3}{l}{**10HMM default element distribution}\\
\noalign{\vskip-9pt}
\multicolumn{3}{l}{***nob: no bremsstrahlung}\\
\end{tabular}
\end{center}

\newcommand{\ispace}{\hskip12pt}
\newcommand{\Fspace}{\hskip9pt}
\newcommand{\colsep}{\hskip9pt}
\begin{planotable}{llllll}
\tablewidth{35pc}
\tablenum{2a}
\tablecaption{}
\tablehead{
	&	&\colhead{pat\_88\_10}	&  &  &\colhead{pat\_88\_tw}}
\startdata
	&$\phi\rightarrow$	&  &	&$\phi\rightarrow$  &
\nl\noalign{\vskip-4pt}
$\theta$&1	&{\small F}\Fspace i\ispace i\ispace i\ispace i\ispace
i\ispace i\ispace i\ispace i\ispace\colsep
&$\theta$&1 	&2\ \ \ 0\ \ \ 0\ \ \ 1\ \ \ 1\ \ \ 1\ \ \ 1\ \ \ 1\ \
\ 0\ \ \ \nl\noalign{\vskip-4pt}
$\downarrow$	&2	&i\ispace i\ispace i\ispace i\ispace i\ispace i\ispace
{\small F}\Fspace i\ispace i\ispace\colsep
&$\downarrow$	&2	&0\ \ \ 0\ \ \ 0\ \ \ 1\ \ \ 1\ \ \ 1\ \ \ 2\ \ \ 1\ \
\ 0\ \ \ \nl\noalign{\vskip-4pt}
	&3	&i\ispace i\ispace i\ispace i\ispace i\ispace i\ispace
i\ispace {\small F}\Fspace i\ispace\colsep
&	&3	&1\ \ \ 1\ \ \ 0\ \ \ 0\ \ \ 0\ \ \ 1\ \ \ 1\ \ \ 2\ \ \ 0\ \ \
\nl\noalign{\vskip-4pt}
	&4	&i\ispace i\ispace  {\small F}\Fspace i\ispace i\ispace
i\ispace i\ispace i\ispace i\ispace\colsep
&	&4	&0\ \ \ 0\ \ \ 2\ \ \ 1\ \ \ 1\ \ \ 1\ \ \ 1\ \ \ 1\ \ \ 1\ \ \
\nl\noalign{\vskip-4pt}
	&5	&i\ispace i\ispace i\ispace i\ispace i\ispace i\ispace
i\ispace {\small F}\Fspace i\ispace\colsep
&	&5	&0\ \ \ 0\ \ \ 0\ \ \ 0\ \ \ 0\ \ \ 1\ \ \ 1\ \ \ 2\ \ \ 1\ \ \
\nl\noalign{\vskip-4pt}
	&6	&i\ispace i\ispace i\ispace i\ispace i\ispace i\ispace
i\ispace i\ispace i\ispace\colsep
&	&6	&1\ \ \ 1\ \ \ 1\ \ \ 0\ \ \ 0\ \ \ 0\ \ \ 0\ \ \ 0\ \ \ 1\ \ \
\nl\noalign{\vskip-4pt}
	&7	&i\ispace i\ispace  {\small F}\Fspace i\ispace i\ispace
i\ispace i\ispace i\ispace i\ispace\colsep
&	&7	&0\ \ \ 0\ \ \ 2\ \ \ 0\ \ \ 0\ \ \ 1\ \ \ 1\ \ \ 1\ \ \ 1\ \ \
\nl\noalign{\vskip-4pt}
	&8	&i\ispace i\ispace i\ispace i\ispace i\ispace  {\small
F}\Fspace i\ispace i\ispace i\ispace\colsep
&	&8	&0\ \ \ 0\ \ \ 0\ \ \ 1\ \ \ 1\ \ \ 2\ \ \ 1\ \ \ 1\ \ \ 1\ \ \
\nl\noalign{\vskip-4pt}
\hline
	&	&{\small f(finger)$=0.097$(F)}
&	&	&{\small $f(0)=0.417$}\nl\noalign{\vskip-4pt}
	&	&{\small f(interstitial)$=0.903$(i)}
&	&	&{\small $f(1)=0.486$}\nl\noalign{\vskip-4pt}
	&	&
&	&	&{\small $f(2)=0.097$}\nl\noalign{\vskip-4pt}
\end{planotable}

\begin{center}
\begin{tabular}{lrl}
\multicolumn{3}{c}{Table 2b.}\\
\hline\hline
\multicolumn{3}{c}{pat\_2424\_tw}\\
\hline
		&$\phi\rightarrow$	&	\cr\noalign{\vskip-7pt}
$\theta$	&1	&\small 2\ \ \ 0\ \ \ 0\ \ \ 0\ \ \ 0\
\ \ 0\ \ \ 0\ \ \ 1\ \ \ 1\ \ \  1\ \ \ 1\ \ \ 1\ \ \ 1\ \ \ 1\ \ \ 1\ \ \ 1\
\ \ 1\ \ \ 1\ \ \ 1\ \ \ 1\ \ \ 1\ \ \ 1\ \ \ 1\ \ \ 1\ \ \ 0\ \ \
\cr\noalign{\vskip-7pt}
$\downarrow$	&2 &\small 1\ \ \ 1\ \ \ 1\ \ \ 1\ \ \ 1\ \ \ 1\ \ \ 1\ \ \ 1\
\ \ 0\ \ \ 0\ \ \ 0\ \ \ 0\ \ \ 0\ \ \ 0\ \ \ 0\ \ \ 1\ \ \ 1\ \ \ 1\ \ \ 1\ \
\ 1\ \ \ 1\ \ \ 1\ \ \ 1\ \ \ 1\ \ \ 0\ \ \ \cr\noalign{\vskip-7pt}
	&3	&\small 1\ \ \ 1\ \ \ 1\ \ \ 1\ \ \ 1\ \ \ 1\ \ \ 1\ \ \ 1\ \
\ 0\ \ \ 0\ \ \ 0\ \ \ 0\ \ \ 0\ \ \ 0\ \ \ 0\ \ \ 1\ \ \ 1\ \ \ 1\ \ \ 1\ \ \
1\ \ \ 1\ \ \ 1\ \ \ 1\ \ \ 1\ \ \ 0\ \ \ \cr\noalign{\vskip-7pt}
	&4	&\small 1\ \ \ 1\ \ \ 1\ \ \ 1\ \ \ 1\ \ \ 1\ \ \ 1\ \ \ 1\ \
\ 0\ \ \ 0\ \ \ 0\ \ \ 0\ \ \ 0\ \ \ 0\ \ \ 0\ \ \ 1\ \ \ 1\ \ \ 1\ \ \ 1\ \ \
1\ \ \ 1\ \ \ 1\ \ \ 1\ \ \ 1\ \ \ 0\ \ \ \cr\noalign{\vskip-7pt}
	&5	&\small 1\ \ \ 1\ \ \ 1\ \ \ 1\ \ \ 1\ \ \ 1\ \ \ 1\ \ \ 1\ \
\ 0\ \ \ 0\ \ \ 0\ \ \ 0\ \ \ 0\ \ \ 0\ \ \ 0\ \ \ 1\ \ \ 1\ \ \ 1\ \ \ 1\ \ \
1\ \ \ 1\ \ \ 1\ \ \ 1\ \ \ 1\ \ \ 0\ \ \ \cr\noalign{\vskip-7pt}
	&6	&\small 1\ \ \ 1\ \ \ 1\ \ \ 1\ \ \ 1\ \ \ 1\ \ \ 1\ \ \ 1\ \
\ 0\ \ \ 0\ \ \ 0\ \ \ 0\ \ \ 0\ \ \ 0\ \ \ 0\ \ \ 1\ \ \ 1\ \ \ 1\ \ \ 1\ \ \
1\ \ \ 1\ \ \ 1\ \ \ 1\ \ \ 1\ \ \ 0\ \ \ \cr\noalign{\vskip-7pt}
	&7	&\small 0\ \ \ 0\ \ \ 0\ \ \ 0\ \ \ 2\ \ \ 0\ \ \ 0\ \ \ 1\ \
\ 1\ \ \ 1\ \ \ 1\ \ \ 1\ \ \ 0\ \ \ 0\ \ \ 0\ \ \ 0\ \ \ 0\ \ \ 0\ \ \ 0\ \ \
1\ \ \ 1\ \ \ 1\ \ \ 1\ \ \ 1\ \ \ 0\ \ \ \cr\noalign{\vskip-7pt}
	&8	&\small 0\ \ \ 0\ \ \ 0\ \ \ 0\ \ \ 0\ \ \ 0\ \ \ 0\ \ \ 1\ \ \ 1\ \ \
1\ \ \ 1\ \ \ 1\ \ \ 0\ \ \ 0\ \ \ 0\ \ \ 0\ \ \ 0\ \ \ 0\ \ \ 0\ \ \ 1\ \ \
1\ \ \ 1\ \ \ 1\ \ \ 1\ \ \ 1\ \ \ \cr\noalign{\vskip-7pt}
&9	&\small 0\ \ \ 0\ \ \ 0\ \ \ 0\ \ \ 0\ \ \ 0\ \ \ 0\ \ \ 1\ \ \ 1\ \ \
1\ \ \ 1\ \ \ 1\ \ \ 0\ \ \ 0\ \ \ 0\ \ \ 0\ \ \ 0\ \ \ 0\ \ \ 0\ \ \ 1\ \ \
1\ \ \ 1\ \ \ 1\ \ \ 1\ \ \ 1\ \ \ \cr\noalign{\vskip-7pt}
&10 	&\small 0\ \ \ 0\ \ \ 0\ \ \ 0\ \ \ 0\ \ \ 0\ \ \ 0\ \ \ 1\ \ \ 1\ \ \
1\ \ \ 1\ \ \ 1\ \ \ 0\ \ \ 0\ \ \ 0\ \ \ 0\ \ \ 0\ \ \ 0\ \ \ 0\ \ \ 1\ \ \
1\ \ \ 1\ \ \ 1\ \ \ 1\ \ \ 1\ \ \ \cr\noalign{\vskip-7pt}
&11	&\small 0\ \ \ 0\ \ \ 0\ \ \ 0\ \ \ 0\ \ \ 0\ \ \ 0\ \ \ 1\ \ \ 1\ \ \
1\ \ \ 1\ \ \ 1\ \ \ 0\ \ \ 0\ \ \ 0\ \ \ 0\ \ \ 0\ \ \ 0\ \ \ 0\ \ \ 1\ \ \
1\ \ \ 1\ \ \ 1\ \ \ 1\ \ \ 1\ \ \ \cr\noalign{\vskip-7pt}
&12	&\small 0\ \ \ 0\ \ \ 0\ \ \ 0\ \ \ 0\ \ \ 0\ \ \ 0\ \ \ 1\ \ \ 1\ \ \ 1\ \
\ 1\ \ \ 1\ \ \ 0\ \ \ 0\ \ \ 0\ \ \ 0\ \ \ 0\ \ \ 0\ \ \ 0\ \ \ 1\ \ \ 1\ \ \
1\ \ \ 1\ \ \ 1\ \ \ 1\ \ \ \cr\noalign{\vskip-7pt}
&13	&\small 0\ \ \ 0\ \ \ 0\ \ \ 0\ \ \ 0\ \ \ 0\ \ \ 0\ \ \ 0\ \ \ 0\ \ \
0\ \ \ 2\ \ \ 0\ \ \ 0\ \ \ 0\ \ \ 0\ \ \ 0\ \ \ 0\ \ \ 0\ \ \ 0\ \ \ 1\ \ \
1\ \ \ 1\ \ \ 1\ \ \ 1\ \ \ 1\ \ \ \cr\noalign{\vskip-7pt}
&14	&\small 0\ \ \ 0\ \ \ 0\ \ \ 0\ \ \ 0\ \ \ 0\ \ \ 0\ \ \ 0\ \ \ 0\ \ \
0\ \ \ 0\ \ \ 0\ \ \ 0\ \ \ 1\ \ \ 1\ \ \ 1\ \ \ 1\ \ \ 1\ \ \ 1\ \ \ 1\ \ \
1\ \ \ 1\ \ \ 1\ \ \ 1\ \ \ 1\ \ \ \cr\noalign{\vskip-7pt}
&15	&\small 0\ \ \ 0\ \ \ 0\ \ \ 0\ \ \ 0\ \ \ 0\ \ \ 0\ \ \ 0\ \ \ 0\ \ \
0\ \ \ 0\ \ \ 0\ \ \ 0\ \ \ 1\ \ \ 1\ \ \ 1\ \ \ 1\ \ \ 1\ \ \ 2\ \ \ 1\ \ \
1\ \ \ 1\ \ \ 1\ \ \ 1\ \ \ 1\ \ \ \cr\noalign{\vskip-7pt}
&16	&\small 1\ \ \ 1\ \ \ 1\ \ \ 1\ \ \ 1\ \ \ 1\ \ \ 1\ \ \ 1\ \ \ 0\ \ \
0\ \ \ 0\ \ \ 0\ \ \ 0\ \ \ 0\ \ \ 1\ \ \ 1\ \ \ 1\ \ \ 1\ \ \ 0\ \ \ 0\ \ \
0\ \ \ 0\ \ \ 0\ \ \ 0\ \ \ 0\ \ \ \cr\noalign{\vskip-7pt}
&17	&\small 1\ \ \ 1\ \ \ 1\ \ \ 1\ \ \ 1\ \ \ 1\ \ \ 1\ \ \ 1\ \ \ 0\ \ \
0\ \ \ 0\ \ \ 0\ \ \ 0\ \ \ 0\ \ \ 1\ \ \ 1\ \ \ 1\ \ \ 1\ \ \ 0\ \ \ 0\ \ \
0\ \ \ 0\ \ \ 0\ \ \ 0\ \ \ 0\ \ \ \cr\noalign{\vskip-7pt}
&18	&\small 1\ \ \ 1\ \ \ 1\ \ \ 1\ \ \ 1\ \ \ 1\ \ \ 1\ \ \ 1\ \ \ 0\ \ \
0\ \ \ 0\ \ \ 0\ \ \ 0\ \ \ 0\ \ \ 1\ \ \ 1\ \ \ 1\ \ \ 1\ \ \ 0\ \ \ 0\ \ \
0\ \ \ 0\ \ \ 0\ \ \ 0\ \ \ 0\ \ \ \cr\noalign{\vskip-7pt}
&19	&\small 1\ \ \ 1\ \ \ 1\ \ \ 1\ \ \ 1\ \ \ 1\ \ \ 1\ \ \ 1\ \ \ 0\ \ \
0\ \ \ 0\ \ \ 0\ \ \ 0\ \ \ 0\ \ \ 1\ \ \ 1\ \ \ 1\ \ \ 1\ \ \ 0\ \ \ 0\ \ \
0\ \ \ 0\ \ \ 0\ \ \ 0\ \ \ 0\ \ \ \cr\noalign{\vskip-7pt}
&20 	&\small 0\ \ \ 0\ \ \ 0\ \ \ 0\ \ \ 0\ \ \ 0\ \ \ 0\ \ \ 1\ \ \ 1\ \ \
1\ \ \ 1\ \ \ 1\ \ \ 1\ \ \ 1\ \ \ 1\ \ \ 1\ \ \ 1\ \ \ 1\ \ \ 1\ \ \ 1\ \ \
1\ \ \ 1\ \ \ 1\ \ \ 1\ \ \ 1\ \ \ \cr\noalign{\vskip-7pt}
&21 	&\small 0\ \ \ 0\ \ \ 0\ \ \ 0\ \ \ 0\ \ \ 0\ \ \ 0\ \ \ 1\ \ \ 1\ \ \
1\ \ \ 1\ \ \ 1\ \ \ 2\ \ \ 1\ \ \ 1\ \ \ 1\ \ \ 1\ \ \ 1\ \ \ 1\ \ \ 1\ \ \
1\ \ \ 1\ \ \ 1\ \ \ 1\ \ \ 1\ \ \ \cr\noalign{\vskip-7pt}
&22	&\small 0\ \ \ 0\ \ \ 0\ \ \ 0\ \ \ 0\ \ \ 0\ \ \ 0\ \ \ 1\ \ \ 1\ \ \
1\ \ \ 1\ \ \ 1\ \ \ 1\ \ \ 1\ \ \ 1\ \ \ 1\ \ \ 1\ \ \ 1\ \ \ 1\ \ \ 1\ \ \
1\ \ \ 1\ \ \ 1\ \ \ 1\ \ \ 1\ \ \ \cr\noalign{\vskip-7pt}
&23	&\small 0\ \ \ 0\ \ \ 0\ \ \ 0\ \ \ 0\ \ \ 0\ \ \ 0\ \ \ 1\ \ \ 1\ \ \
1\ \ \ 1\ \ \ 1\ \ \ 1\ \ \ 1\ \ \ 1\ \ \ 1\ \ \ 1\ \ \ 1\ \ \ 1\ \ \ 1\ \ \
1\ \ \ 1\ \ \ 1\ \ \ 1\ \ \ 1\ \ \ \cr\noalign{\vskip-7pt}
&24	&\small 0\ \ \ 0\ \ \ 0\ \ \ 0\ \ \ 0\ \ \ 0\ \ \ 0\ \ \ 1\ \ \ 1\ \ \
1\ \ \ 1\ \ \ 1\ \ \ 1\ \ \ 1\ \ \ 1\ \ \ 1\ \ \ 1\ \ \ 1\ \ \ 1\ \ \ 1\ \ \
1\ \ \ 1\ \ \ 1\ \ \ 1\ \ \ 1\ \ \ \cr
\noalign{\hrule}
&2424s:	&$f(0)=0.437$\cr
&	&$f(1)=0.555$\cr
&	&$f(2)=0.00833$\cr
\noalign{\hrule}
\end{tabular}
\end{center}

\begin{planotable}{rlllll}
\tablewidth{40pc}
\tablenum{3}
\tablecaption{}
\tablehead{
\multicolumn{6}{c}{Terminal Letter Model Designators\phantom{abcdefghij}}}
\startdata
e:	&\multicolumn{5}{l}{Flat compositions to 3000 km/s, conserving mass}\nl
f:	&\multicolumn{5}{l}{Flat compositions at central $Y_i$, conserving
	mass}\nl
c:	&\multicolumn{5}{l}{Centrally concentrated $^{56}$Ni, out to only
	7$\mdot$}\nl
sph:	&\multicolumn{5}{l}{Spherical distribution of elements, no clumping or
	heterogeneity}\nl
\hline
\noalign{\vspace{3pt}}
\hline
\noalign{\vspace{3pt}}
\noalign{\centerline{\bf Two-Phase}}
\noalign{\vspace{3pt}}
\hline
\noalign{\vspace{3pt}}
88h:	&\multicolumn{5}{l}{\small H/He in fingers, $Z>2$ between fingers,
pat\_88\_10, finger volume fraction$=$0.097}\nl
2424g:	&\multicolumn{5}{l}{\small Ni and H in space between fingers, finger
volume fraction$=$0.03, pat\_2424\_30}\nl
2424h:	&\multicolumn{5}{l}{\small H and He between fingers, $Z>2$ in
fingers, finger volume fraction$=$0.0083, pat\_2424\_100}\nl
2424h$'$:&\multicolumn{5}{l}{\small Ni and H in space between fingers, finger
volume fraction$=$0.0083, pat\_2424\_100}\nl
2424k:	&\multicolumn{5}{l}{All $Z>2$ in fingers, H/He in between,
	pat\_2424\_100}\nl
\hline
\noalign{\vspace{3pt}}
\hline
\noalign{\vspace{3pt}}
\noalign{\centerline{\bf Three-Phase}}
\noalign{\vspace{3pt}}
\hline
\noalign{\vspace{3pt}}
{\small Composition:}
&\multicolumn{1}{c}{0}
&\multicolumn{1}{c}{1}
&\multicolumn{1}{c}{2}
&	&	\nl
\hline
\noalign{\vspace{3pt}}
&\multicolumn{1}{c}{{\small $f(26>Z>2)$}}
&\multicolumn{1}{c}{$f({\rm H/He})$}
&\multicolumn{1}{c}{$f({\rm Ni})$}
&\multicolumn{1}{c}{Pattern}
&\multicolumn{1}{l}{\qquad Comments}\nl
\hline
88t:	&\qquad0.819 	&\quad0.083 	&\qquad 0.097	&pat\_88\_tw2
	&clumping to 5000 km/s\nl
88u:	&\qquad0.097	&\quad0.72	&\qquad 0.18
	&\multicolumn{1}{c}{$\dag$}	&clumping to 5000 km/s\nl
88r:	&\qquad0.819	&\quad0.083	&\qquad 0.097	&pat\_88\_tw2
	&clumping to surface\nl
88v:	&\qquad0.097	&\quad0.79	&\qquad 0.11
	&\multicolumn{1}{c}{$\dag$}	&clumping to 5000 km/s\nl
88p:	&\qquad0.097	&\quad0.486	&\qquad 0.417
	&\multicolumn{1}{c}{$\dag$}	&clumping to surface\nl
88q:	&\qquad0.417	&\quad0.486	&\qquad 0.097	&pat\_88\_tw
	&clumping to surface\nl
2424s:	&\qquad0.44	&\quad0.56	&\qquad 0.0083	&pat\_2424\_tw
	&clumping to 5000 km/s\nl
88s:	&\qquad0.417	&\quad0.486	&\qquad 0.097	&pat\_88\_tw
	&clumping to 5000 km/s\nl
\hline
\multicolumn{6}{l}{$^{\dag}$ Started with a 2-phase pattern pat\_88\_10, and
then third phase added at random.}\nl
\end{planotable}